\documentclass[10pt,conference]{IEEEtran}
\IEEEoverridecommandlockouts
\usepackage{cite}
\usepackage{amsmath,amssymb,amsfonts}
\usepackage{algorithmic}
\usepackage{graphicx}
\usepackage{textcomp}
\usepackage{xcolor}
\usepackage{multirow}
\usepackage{mathtools}
\DeclarePairedDelimiter\ceil{\lceil}{\rceil}
\usepackage{makecell}
\usepackage{caption}
\usepackage{subcaption}
\newtheorem{insight}{Insight}
\def\BibTeX{{\rm B\kern-.05em{\sc i\kern-.025em b}\kern-.08em
    T\kern-.1667em\lower.7ex\hbox{E}\kern-.125emX}}
\newcommand{\ignore}[1]{}

\begin{document}

\title{A Case Study on Using Deep Learning for Network Intrusion Detection}

\author{\IEEEauthorblockN{Gabriel C. Fern\'andez}
\IEEEauthorblockA{Department of Computer Science \\
University of Texas at San Antonio
% San Antonio, TX, USA \\
%qhu044@my.utsa.edu
}
\and
\IEEEauthorblockN{Shouhuai Xu}
\IEEEauthorblockA{Department of Computer Science \\
University of Texas at San Antonio
% San Antonio, TX, USA \\
%shouhuai.xu@utsa.edu
}
}

\maketitle

%\tableofcontents

\begin{abstract}
Deep Learning has been very successful in many application domains. However, its usefulness in the context of network intrusion detection has not been systematically investigated. In this paper, we report a case study on using deep learning for both supervised network intrusion detection and unsupervised network anomaly detection. We show that Deep Neural Networks (DNNs) can outperform other machine learning based intrusion detection systems, while being robust 
%against
in the presence of dynamic IP addresses.
%the Dynamic Host Configuration Protocol (DHCP). 
%when including IP address as a feature. 
We also show that Autoencoders can be effective for network anomaly detection. 
%In this paper we report a case study on using deep learning for network intrusion detection.
\end{abstract}

\begin{IEEEkeywords}
Network Intrusion Detection, Deep Learning, Deep Neural Network, Autoencoder, Anomaly Detection
%Supervised Learning, Unsupervised Learning
\end{IEEEkeywords}

\section{Introduction}

%\subsection{Research motivation}
%explain why we are doing this research and why it is important
As the scale of cyber attacks and volume of network data increases exponentially, organizations must 
%develop new ways of keeping their networks and data secure from the dynamic nature of evolving threat actors.
continually adapt against the dynamic nature of evolving cyber threat actors.
With more security tools and sensors being deployed in modern enterprise networks, the number of security events being generated continues to increase, making it more challenging to 
%find the needle in the haystack.
detect malicious activities.
%New techniques must be developed to augment human analysts when it comes to the 
Organizations must adopt new techniques to augment human analysts in monitoring, preventing, detecting, and responding to cybersecurity events and potential attacks.
%As the adversary continues to evolve, so must the defender.
Machine Learning has been deemed by many as a game changer in cyber defense. 
However, the usefulness of Deep Learning in the context of Network Intrusion Detection Systems (NIDSs) has not been systematically understood, despite its tremendous success in other application domains (e.g., image recognition). 

\subsection{Our contributions}
%(NOTE: This text is from the abstract of the Thesis. Perhaps this can be moved to the abstract of this paper, and update this section with more specifics/alternate information)
%The focus for this paper is on classifying network traffic flows as benign or malicious.
The contribution of this work is two-fold. 
First, we propose using a feedforward fully connected Deep Neural Network (DNN) to train a NIDS via supervised learning.
We also propose using an autoencoder to detect and classify attack traffic via unsupervised learning in the absence of labeled malicious traffic.
Second, we evaluate these models using two recent network intrusion detection datasets with known ground truth of malicious vs. benign traffic. We show (i) DNN outperforms other machine learning based network intrusion detection systems; (ii) DNN is robust 
%against
in the presence of dynamic IP addresses assigned by the Dynamic Host Configuration Protocol (DHCP), which is important when we need to use IP addresses as features in training DNNs; and (iii) autoencoder is effective for anomaly detection.

\subsection{Related work}

Intrusion detection can be host-based or network-based, with this paper being in the latter category. There are multiple approaches to network-based intrusion detection. The idea of {\em anomaly detection} can be traced back as early as the 19th century with origins in the statistics community \cite{edgeworth1887xli}.
The study of intrusion detection for cybersecurity is introduced in 1987 \cite{Denning:1987fa}.
Fiore et al. \cite{Fiore13} explored the use of a semi-supervised model for network intrusion detection, using a Discriminative Restricted Boltzmann Machine. 
%They train a classifier on the ``normal'' traffic only, with the goal of detecting anomalous behaviors that may evolve over time.
%They argue that it is often difficult to train a model with anomalous training samples \textit{a priori}.
However, their study is based on the KDD 99 dataset \cite{Cup1999data}, which is outdated now \cite{Sommer:2010jv}.
%For example, Sommer et al. \cite{Sommer:2010jv} suggest that the KDD 99 dataset may be better suited for only providing additional validation and cross-checking of a novel technique.
% Niyaz et. al. [CITE] explore a deep learning based technique, however they use the NSL-KDD dataset, which is also based off KDD 99.

%A primary and ongoing challenge in the field of network intrusion detection is the lack of publicly available, labeled datasets that can be used for effective testing, evaluation, and comparison of techniques \cite{Niyaz2015,Shiravi:2012}.
In addition to the KDD 99 dataset, 
%Granted, publicly labeled datasets are available, 
there are other datasets, such as: the CAIDA dataset \cite{Caida2011}, the DARPA/Lincoln Lab packet trace \cite{Lippmann1999results
, Lippmann20001999
}, and the Lawrence Berkeley National Laboratory (LBNL) and ICSI Enterprise Tracing Project \cite{LBNL:2005}.
However, comparative studies of these datasets \cite{Sharafaldin:2018jr,Shiravi:2012} 
%performed a comparative study of all of these datasets %since 1998 
found that some are 
%mostly anonymized and do not contain valuable payload information, making them less useful for research purposes 
outdated and unreliable because they lack a diversity of traffic and volumes, some lack a variety of attacks, some are anonymized and lack valuable payload information, and some lack feature set and metadata.
%\cite{Sharafaldin:2018jr, Shiravi:2012}.
% While these datasets have proven useful, there are some arguments as to the validity of using them in present day research --- they may be better suited for the purposes of providing additional validation and cross-checking of a novel technique \cite{Sommer:2010jv}.
% As newer intrusion detection benchmark datasets have been introduced (two of which will be utilized in this paper -- ISCX IDS 2012 and CIC IDS 2017), there is now a more useful facility for which methods and approaches of NIDS can be compared and evaluated.

The present paper focuses on using newer datasets that have recently become available to the research community.
%, specifically the UNB ISCX IDS 2012 and UNB CIC IDS 2017 datasets.
Not only do these newer datasets contain modern-day attacks, they are created in such a way as to follow established guidelines of reliable intrusion detection datasets (in terms of realism, evaluation capabilities, total capture, completeness, and malicious activity) \cite{Shiravi:2012}.
%The CIC IDS 2017 dataset is even stronger in that it adheres to all eleven criteria that have been identified as necessary for building a reliable benchmark dataset, whereas previous datasets have not \cite{Sharafaldin:River2017}.
There are a number of other studies that use these datasets for evaluation in their work.
However, many evaluating the ISCX IDS 2012 dataset \cite{Shiravi:2012} use only a subset of the data, and vary in their ways for generating the ground truth 
%(due to nature of ISCX IDS 2012 dataset)
%yassin2013anomaly
%kumar2013design
\cite{yassin2013anomaly,ammar2015decision,folino2016distributed,tan2015detection,atli2017anomaly}.
Studies conducted with the CIC IDS 2017 dataset \cite{Shiravi:2012} have used other types of machine learning techniques than Deep Learning \cite{ahmim2018novel, de2018experimental, adnan2017forest, ibarguren2015coverage, chang2001libsvm, huhn2009furia, aksu2018intrusion}.
%as compared further in this work in Table \ref{table:cicids-result-comparison}
More recent studies have begun to use Deep Learning with the CIC IDS 2017 dataset; however, some only use a subset of the data for detecting one type of attack (e.g., port scan, DDoS) \cite{aksu2018detecting,jiang2018aldd,ustebay2018intrusion} or generate their own flows instead of using the ground truth flows \cite{pektacsdeep}.
The present study differs in that it evaluates both supervised and unsupervised deep learning approaches across the full spectrum of attacks, while using the entirety of the datasets as well as the ground truth provided. 
%by the benchmark datasets.
%for robust comparison against other approaches.
%{\color{red}To the best of our knowledge, this is the first paper that use deep learning methods for network intrusion detection while tested by these new datasets.}
%\footnote{need to cite other papers, if any, that analyze these two datasets and highlight the difference between their papers and ours}

Intrusion detection is an important field of cybersecurity data analytics \cite{XuTIFSDataBreach2018,VulDeePecker,XuVulPeckerACSAC2016,XuIEEECNS2014,XuCODASPY13,XuIEEETIFS2013,XuIEEETIFS2015,XuPLoSOne2015,FickeIEEEMilcom2018}, which is one pillar underlying the Cybersecurity Dynamics framework \cite{XuCybersecurityDynamicsHotSoS2014,XuBookChapterCD2018} that aims to model and quantify cybersecurity from a holistic perspective. 
The other two pillars are known as first-principle modeling and analysis \cite{XuTAAS2012,XuTDSC2011,XuGameSec13,XuTAAS2014,XuHotSOS14-MTD,XuIEEETNSE2018,XuIEEEACMToN2019} and cybersecurity metrics \cite{Pendleton16,XuSTRAM2018manuscript,XuIEEETIFS2018-groundtruth,XuAgility2019}.
%There are many sub-fields in cybersecurity data analytics. The present paper falls into prescriptive cybersecurity data analytics aim to detect network attacks, which are then treated properly by the defender; therefore, this sub-field of cybersecurity data analytics may be called {\em reactive cybersecurity data analytics}.

% discuss related prior studies ... explain how they are related to ours ... and what are the differences between these studies and ours (i.e., why our paper deserves to be published)

\ignore{

\begin{figure}[htbp]
\centerline{\includegraphics[scale=0.17]{figures/lcd-framework.png}}
\caption{Cybersecurity Dynamics Framework}
\label{fig:lcd-framework}
\end{figure}

}

%\footnote{given the limited space, focus on discussing prior studies on NIDS with or without using deep learning; clarify the differences between the present paper and prior studies in terms of problems and/or approaches}

\smallskip

The rest of the paper is organized as follows. Section \ref{sec:preliminaries} reviews DNNs and Autoencoders. 
Section \ref{sec:case-study} presents the case study.
Section \ref{sec:limitations} discusses the limitations of the present study.
Section \ref{sec:conclusion} concludes the paper.

\section{Preliminaries}
\label{sec:preliminaries}

% use at most half-page to review DNN and Autoencoder so that the paper is self-contained ... explain why we consider these two but not others ...

% Deep learning, a subfield of machine learning, excels in generalizing to new examples when the data is complex in nature and contains a high level of dimensionality \cite{Goodfellow-et-al-2016}. 
% In addition, deep learning enables the scalable training of nonlinear models on large datasets \cite{Goodfellow-et-al-2016}.
% This is important in the domain of network intrusion detection because not only is it dealing with a large amount of data, but the model generated by the deep learning system will need to be capable of generalizing to new forms of attacks not specifically represented in the currently available labeled data.

\begin{figure}[!htbp]
\centerline{\includegraphics[scale=0.28]{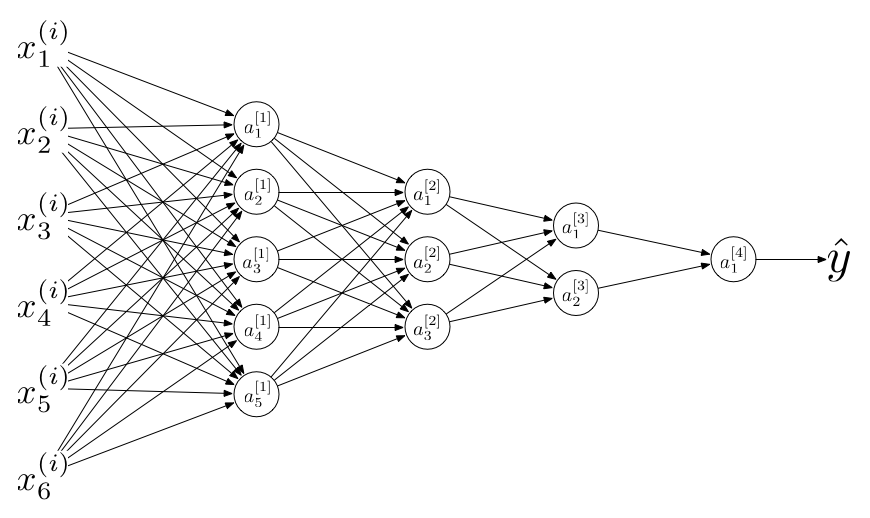}}
\caption{Deep neural network representation}
\label{fig:dnn}
\end{figure}

DNNs are a powerful mechanism for supervised learning.
They can represent functions of increasing complexity, by inclusion of more layers and more units per layer in a neural network \cite{Goodfellow-et-al-2016}.
%The underlying technology and algorithms in deep learning are based on the utilization of neural network architectures consisting of multiple layers of neurons.
In the context of NIDSs, DNNs can be used to discover patterns of benign and malicious traffic hidden within large amounts of structured data.
Figure \ref{fig:dnn} is an example of a standard Deep Learning representation, where nodes represent inputs, edges represent weights, superscript $(i)$ denotes the $i$th training example, and superscript $[l]$ denotes the $l$th layer.
%DNNs are one type of deep learning architecture, in addition to recurrent deep neural networks (RNNs) and convolutional deep neural networks (CNNs).
Our case study focuses on DNNs because they can cope with tabular data 
%for which there exists lots of examples, 
and categorical variables of high cardinality, which are exhibited by the datasets we analyze.

%Future studies should be performed utilizing RNNs, as they are well suited for leveraging sequence and time domain information.
%CNNs can be useful when payload information is taken into consideration, as was done in \cite{lotfollahi2017deep} where they used deep learning for traffic classification.
%This paper focuses on the use of flow level data, which omits the full packet and instead leverages packet metadata and statistics -- covering more breadth while utilizing less space.

\begin{figure}[!htbp]
\centerline{\includegraphics[scale=0.24]{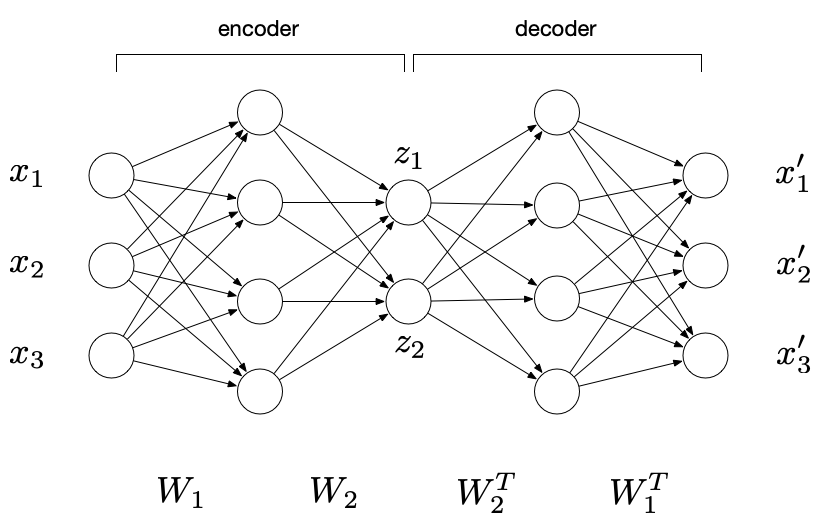}}
\caption{Example Autoencoder neural network architecture}
\label{fig:autoencoder}
\end{figure}

Autoencoder is another type of neural network and is trained in such a way that it aims to copy its input to its output,
% Autoencoders are useful in cases where there are lots of examples of what normal data looks like, yet it is difficult to explain what represents anomalous activity.
% For this reason, autoencoders can be powerful when used in anomaly detection systems.
namely 
%Autoencoder's main utility is 
aiming to find a lower dimensional, latent space representation of the input data  \cite{Goodfellow-et-al-2016}.
Unlike other popular dimensionality reduction techniques such as Principle Component Analysis (PCA), it achieves its goal in a non-linear fashion.
%In the context of NIDs, a compressed, latent space representation of benign traffic can be learned via an autoencoder, whereby future unseen traffic is reconstructed based on this learned representation.
%If there is a high reconstruction error, the traffic can be flagged as anomalous.
%Therefore, autoencoders can be used to power an anomaly detection system.
% The key idea behind autoencoders is to take an input vector and map it to a latent space (encode) of lower dimensionality, then from that latent space, map it back to an output (decode) using the low dimensional latent space as input.
Figure \ref{fig:autoencoder} shows an example standard Autoencoder, where the number of input neurons is equal to the number of output neurons.
We choose Autoencoder for our case study on anomaly detection because of its usefulness given lots of normal data, 
%and when it may be difficult to explain what represents anomalous data.
and its applicability to situations where it may be difficult to explain what represents anomalous data.

\section{Case Study}
\label{sec:case-study}

\subsection{Methodology}
Deep Learning 
%enables the scalable training of nonlinear models on large datasets \cite{Goodfellow-et-al-2016}.
excels when there is a large amount of training data \cite{Goodfellow-et-al-2016}.
%According to \cite{Goodfellow-et-al-2016}, the general rule of thumb as of 2016 is that supervised deep learning algorithms will achieve good performance with at least 5,000 labeled examples per category.
%They will also exceed human performance when they are trained with a dataset that has at least 10 million labeled examples.
This suggests that we use the newer datasets for our case study:
the ISCX IDS 2012 dataset 
\cite{Shiravi:2012} has over 2.54M examples (including 2.47M benign ones and 68,910 malicious ones); the CIC IDS 2017 dataset \cite{Sharafaldin:2018jr} has over 2.83M examples (including 2.27M benign ones and 557,646 malicious ones).
%These datasets
%, while not quite at the 10M mark, 
%are much larger than any IDS datasets used in the past
%, and are well-suited for experimenting to determine the effectiveness of deep learning architectures as applied to NIDS.
%and can be used to experiment and determine the effectiveness of deep learning architectures and algorithms as applied to the domain of network intrusion detection.
In contrast, older datasets are often small (e.g., the KDD 99 dataset \cite{Cup1999data} has only 148,517 flows, including 77,054 benign ones and 71,463 malicious ones).

We choose DNNs because they can cope with tabular data that contains categorical variables of high cardinality, which are exhibited by the two newer datasets we use.
A key issue is to cope with categorical features of high cardinality \cite{guo2016entity}.
The idea is to use \textit{entity embedding}
%, which is used to leverage the power of deep learning when working with tabular data \cite{thomas2018fastai}.
to
%Embeddings 
map categorical features of high cardinality to low-dimensional real vectors in such a way that similar values remain close to each other \cite{google2019mlcrashcourse,thomas2018fastai}.

%One of the hallmarks of deep learning is its ability to take complex raw data and create higher-order features automatically in order to make the task of generating a classification or regression output simpler \cite{Patterson:2017:DLP:3169957}.
%The larger datasets used in this work are suited well for evaluation of deep learning as applied to NIDS.
%to help find malicious activity buried within the haystack of flow records, and are similar to the network traffic and log data available on an enterprise network.

We choose Autoencoders because they are useful when there are lots of examples of normal data, while it may be difficult to explain what represents anomalous data 
\cite{Shiravi:2012}.
%For this reason, Autoencoders can be powerful when used in anomaly detection systems.
Autoencoders learn a compressed representation of the input data, meaning that its output is a reconstruction of the input data in a certain form.
By minimizing the error of reconstructing the normal input (i.e., benign flows), Autoencoders learn to modify the weights for reconstructing the input. 
When an Autoencoder encounters a malicious flow, the reconstruction error would be high (in comparison to reconstructing a normal flow).
%, as the malicious flows are anomalous when compared to the benign flows in terms of encoding and decoding.

% describes the methodology for our case study ...explain why we consider both deep neural networks and autoencoders ...

\subsection{Data description}

The ISCX IDS 2012 dataset \cite{Shiravi:2012}
%, who 
%with the University of New Brunswick's Information Security Center of Excellence (ISCX) 
%developed a systematic approach to generating datasets for network intrusion detection purposes.
was created
%the dataset 
by 
%statistically 
modeling a given network environment with a testbed and then using agents to perform attacks on the testbed network.
%; see  \cite{Shiravi:2012} for details. 
%, which consists of network traffic generated on a testbed environment in their laboratory.
When compared with the 
%widely used, but now 
outdated datasets \cite{Lippmann1999results, Lippmann20001999,Cup1999data}, this dataset can be characterized as follows \cite{Shiravi:2012}:
realistic network configuration because of the real testbed; realistic traffic because of the real attacks/exploits;
%(to the extent of specified profiles); 
labeled ground truth of benign and malicious traffic; total capture of communications; 
%	\item{Complete capture}
and diverse attack scenarios.
%are involved.
%{\color{red}The dataset is in raw PCAP files that are formatted as a custom XML file.}
This dataset is provided in PCAP as well as a custom XML file of {\em network flows} created with the IBM QRadar appliance; the XML flow file contains ground truth labels.
Recall that a {\em network flow} is assembled from a number of IP packets and consists of source and destination IP addresses, source and destination port numbers, and
%the source interface, 
protocol.
%, and the number of bytes transmitted. 
Moreover, flows are often used as a unit for detecting attacks, which is our focus in the present study (another unit is IP packet). 
%This dataset has some limitations: (i) the ground truth labels provided are only binary (e.g., malicious or benign); and
%(ii) lacks a CSV version of flow records which can be consistently used by other studies;
%{\color{red}it only comes with attack vs. benign traffic, meaning that it can only be used for binary classification};
%as opposed to just a binary classifier in the case of the labeled ISCX IDS 2012 dataset.
%(ii) it contains no traffic of the HTTPS protocol.
%s despite that 
%, which is important since 
%over $70\%$ of traffic on the Internet is now traversing the HTTPS protocol\cite{Sharafaldin:2018jr}; and
%(iv) the distribution of the simulated attacks is not based on real world statistics \cite{Sharafaldin:2018jr, Shiravi:2012}. 
Table \ref{table:UNB_ISCX_2012} provides an overview of this dataset. 
%%%%%%%%%%%%%%%%%%%%%%%%%%%%%%%%
%%%% ISCX 2012 Dataset Table %%%
%%%%%%%%%%%%%%%%%%%%%%%%%%%%%%%%
\begin{table}[!htbp]
\caption{Overview of the ISCX IDS 2012 dataset, where 
%``date'' is the day on which the network traffic is captured, 
%``\# of flows'' is the total number of network flows on a given day, 
``\# of attacks'' is the subset of flows that contain an attack.
%, and ``description'' indicates the type of attack performed.
}
%{\color{red}``date'' means ... explain the meaning of each column'' ... how many different kinds of attacks are involved ..}
\begin{center}
\begin{tabular}{|c|c|c|l|}
\hline 
\textbf{Date} & \textbf{\# of Flows}& \textbf{\# of Attacks}& \textbf{Description} \\
\hline
6/11/2012& 474,278& 0& Benign network activities\\
\hline
6/12/2012& 133,193& 2,086& Brute-force against SSH\\
\hline
6/13/2012& 275,528& 20,358& Infiltrations internally\\
\hline
6/14/2012& 171,380& 3,776& HTTP DoS attacks\\
\hline
6/15/2012& 571,698& 37,460& DDoS using IRC bots\\
\hline
6/16/2012& 522,263& 11& Brute-force against SSH\\
\hline
6/17/2012& 397,595& 5,219& Brute-force against SSH\\
\hline
\textbf{Total}& \textbf{2,545,935}& \textbf{68,910}& \textbf{2.71\% malicious} \\
\hline
%\multicolumn{4}{l}{$^{\mathrm{a}}$Sample of a Table footnote.}
\end{tabular}
\label{table:UNB_ISCX_2012}
\end{center}
\end{table}

Table \ref{table:iscx2012-features} summarizes the 14 features that can be extracted from the labeled XML file of network flows.
%from the ISCX IDS 2012 dataset. 
%We observe that there are seven categorical features, with a total number of unique possible values being $2,478 + 34,552 + 64,482 + 24,238 + 107 + 4 + 6 = \textbf{125,867}$. This is important because, as shown below, we need to represent these categorical feature values appropriately for training DNNs.

%as follows: Source IP, Destination IP, Source Port, Destination Port, App Name, Direction, Protocol, Duration, Total Source Bytes, Total Destination Bytes, Total Bytes, Total Source Packets, Total Destination Packets, Total Packets.
\begin{table}[!htbp]
\caption{Description of the 14 features of the ISCX IDS 2012 dataset, where ``uniques'' means the number of possible values of a categorical feature.}
\centering
\label{table:iscx2012-features}
\begin{tabular}{|c|c|l|c|c|}
\hline
\textbf{No.} & \textbf{Feature} & \textbf{Description} & \textbf{Type} & \textbf{Uniques} \\ \hline
1 &SrcIP    &  Source IP address & Categorical & 2,478 \\ \hline
2 &DstIP    &  Dest. IP address & Categorical & 34,552  \\ \hline
3 &SrcPort  &  Source port & Categorical & 64,482 \\ \hline %For ICMP flow, ICMP type.
4 &DstPort &  Dest. port & Categorical & 24,238\\ \hline  %For ICMP flows, ICMP type.
5 &AppName & Application name & Categorical & 107 \\ \hline
6 &Direction & Direction of flow & Categorical & 4 \\ \hline
7 &Protocol &  IP protocol & Categorical & 6 \\ \hline
8 &Duration &  Flow duration & Continuous & N/A\\ \hline
 %Start Time &  Flow start time in ISO 8601 format, with milliseconds\\ \hline
9 &TotalSrcBytes & Total source bytes & Continuous & N/A\\ \hline
10 &TotalDstBytes & Total dest. bytes & Continuous & N/A\\ \hline
11&TotalBytes & Total bytes & Continuous & N/A\\ \hline
12 &TotalSrcPkts & Total source packets & Continuous & N/A\\ \hline
13 &TotalDstPkts & Total dest. packets & Continuous & N/A\\ \hline
14 &TotalPkts & Total packets & Continuous & N/A\\ \hline
 %\# of Packets &  Forward packet count\\ \hline
 %\# of Bytes & Forward octet count \\ \hline
\end{tabular}
\end{table}

The CIC IDS 2017 dataset \cite{Sharafaldin:2018jr} 
%It comes from a collaboration between the Canadian Institute for Cybersecurity (CIC) and University of New Brunswick's Information Security Center of Excellence (ISCX).
% The dataset was created in 2017 and published for the research community to use in 2018.
improves the ISCX IDS 2012 dataset by containing, along with benign traffic, attack traffic from seven different kinds of attacks (i.e., brute-force against the SSH and Web, Heartbleed, botnet, denial of service (DoS), distributed denial of service (DDoS), cross-site scripting (XSS) and SQL injection attacks against websites, and infiltration). 
%{\color{red}...which 7 attacks}
This dataset includes not only the raw PCAP data, but also pre-processed network flow data from the PCAP data (processed using the CICFlowMeter tool \cite{LashkariDraper-Gil:icissp17}).
%, which is a network traffic flow generator 
%written in Java 
%that takes raw PCAP as input and generates bidirectional flows where the first packet determines the forward (source to destination) and backward (destination to source) directions.
%It generates 77 statistical features 
%(the features available for CIC IDS 2017 specifically are detailed in Table \ref{tab:cicids2017-features}), 
%such as duration, number of packets, number of bytes, length of packets, which are also calculated separately in the forward and backwards directions.
%Along with these statistical features for each flow, CICFlowMeter provides the corresponding source/destination IP address and port, protocol number, timestamp, label, and a unique FlowID for each flow record.
This pre-processed network flow data is provided as CSV files that can be fed into the machine learning pipeline. The pre-processed network flow data has 83 columns (e.g., duration, number of packets, number of bytes, length of packets) that can be used as features, plus one label column and one flow ID column.
%, , which is advantageous for evaluating various features within deep learning approaches for NIDSs.
Since seven different kinds of attacks are contained in this dataset, we can conduct multiclass classification research. 
Table \ref{table:cicids2017-dataset-overview}
shows a summary of this dataset.

%%%%%%%%%%%%%%%%%%%%%%%%%%%%%%%%%
%%%% CICIDS2017 Dataset Table %%%
%%%%%%%%%%%%%%%%%%%%%%%%%%%%%%%%%
\begin{table}[!htbp]
\caption{Overview of the CIC IDS 2017 dataset, where the columns have the same meanings as in Table \ref{table:UNB_ISCX_2012}.}
\label{table:cicids2017-dataset-overview}
\begin{center}
\begin{tabular}{|c|c|c|l|}
\hline 
\textbf{Date} & \textbf{\# of Flows}& \textbf{\# of Attacks}& \textbf{Description} \\
\hline
Monday & 529,918 & 0& Normal activities\\
\hline
\multirow{2}{*}{Tuesday} & \multirow{2}{*}{445,909} & 7,938 & FTP-Patator \\
\cline{3-4}
 & & 5,897 & SSH-Patator \\

\hline
\multirow{5}{*}{Wednesday} & \multirow{5}{*}{692,703} & 5,796 & DoS slowloris\\
\cline{3-4}
& & 5,499 & DoS Slowhttptest \\
\cline{3-4}
& & 231,073 & DoS Hulk \\
\cline{3-4}
& & 10,293 & Dos GoldenEye \\
\cline{3-4}
& & 11 & Heartbleed \\

\hline
\multirow{3}{*}{Thursday AM} & \multirow{3}{*}{170,366} & 1507 & Web - Brute Force\\
\cline{3-4}
& & 652 &  Web - XSS\\
\cline{3-4}
& & 21 & Web - SQL Injection \\

\hline
Thursday PM & 288,602 & 36& Infiltration\\
\hline
Friday AM & 191,033 & 1966& Bot\\
\hline
Friday PM 1 & 286,467 & 158,930& PortScan\\
\hline
Friday PM 2 & 225,745 & 128,027& DDoS\\
\hline
\textbf{Total}& \textbf{2,830,743} & \textbf{557,646} & \textbf{19.70\% malicious} \\
\hline
%\multicolumn{4}{l}{$^{\mathrm{a}}$Sample of a Table footnote.}
\end{tabular}
\label{table:CICIDS2017-overview}
\end{center}
\end{table}

Table \ref{table:cicids2017-features} highlights some of the 74 features that were ``useable'' from the CIC IDS 2017 dataset, while noting that among the other 85-74=11 features, eight continuous features contain no variability or missing values and therefore are discarded, 
%{\color{red}These eight features include} 
and the remaining three are FlowID, the timestamp, and the label (used for the predicted class).
%We observe that there are five categorical features, with the total number of unique possible values being $17,002 + 19,112 + 64,638 + 53,791 + 3 = \textbf{154,546}$.

\begin{table}[!htbp]
\caption{Description of some of the 74 features of the CIC IDS 2017 dataset, where the columns have the same meanings as in Table \ref{table:iscx2012-features}.}
\centering
\label{table:cicids2017-features}
\begin{tabular}{|c|c|l|c|c|}
\hline
\textbf{No.} & \textbf{Feature} & \textbf{Description} & \textbf{Type} & \textbf{Uniques} \\ \hline
1 &SrcIP    &  Source IP address & Categorical & 17,002 \\ \hline
2&DstIP    &  Dest. IP address & Categorical & 19,112  \\ \hline
3 &SrcPort  &  Source port & Categorical & 64,638 \\ \hline 
4 &DstPort &  Dest. port & Categorical & 53,791\\ \hline  
5 &Protocol &  IP protocol & Categorical & 3 \\ \hline
6 &Duration &  Flow duration & Continuous & N/A\\ \hline
7 &total\_fpackets & \makecell[l]{Total num. \\ forward packets} & Continuous & N/A\\ \hline
8 &total\_bpackets & \makecell[l]{Total num. \\ backward packets} & Continuous & N/A\\ \hline
9 &total\_fpktl & \makecell[l]{Total size of \\ forward packets} & Continuous & N/A\\ \hline
10 &total\_bpackets & \makecell[l]{Total size of \\ backward packets} & Continuous & N/A\\ \hline
$\vdots$ &$\vdots$ & $\vdots$  & $\vdots$ & $\vdots$ \\ \hline
%69 &max\_active & \makecell[l]{Max time flow\\active before idle} & Continuous& N/A\\ \hline
70 &std\_active & \makecell[l]{Std. dev time flow \\ active before idle} & Continuous & N/A\\ \hline
71 &min\_idle & \makecell[l]{Min time flow \\ idle before active} & Continuous & N/A\\ \hline
72 &mean\_idle & \makecell[l]{Mean time flow \\ idle before active} & Continuous & N/A\\ \hline
73 &max\_idle & \makecell[l]{Max time flow \\ idle before active} & Continuous & N/A\\ \hline
74 &std\_idle & \makecell[l]{Std. dev time flow \\ idle before active} & Continuous & N/A\\ \hline
 %\# of Packets &  Forward packet count\\ \hline
 %\# of Bytes & Forward octet count \\ \hline
\end{tabular}
\vspace{-3mm}
\end{table}

\subsection{Using DNNs for network intrusion detection}

\subsubsection{Pre-processing}
We propose formatting a dataset (more specifically, network flows) in such a way that can be input into a DNN.
%Network traffic data is commonly captured in the format of raw packet capture (PCAP) files, which contain the full TCP/IP packet data transmitted or received on a given network device. While full packet capture information can be useful for certain use cases, it does bring with it a high cost in terms of space. 
%An alternative (or complement) to PCAP data is \textit{flow records}, which serves to summarize the PCAP data in terms of higher level network flows.
%They can have many more fields extracted from the PCAP based on the configuration of software used for converting PCAP to NetFlow records. 
%An emerging IETF standard called IPFIX, short for IP Flow Information Export, is similar to the NetFlow standard, but allows for additional information that is normally sent to syslog, as well as variable length fields for collecting information such as URLs, messages, HTTP hosts, and more \cite{rfc7011,pcwdld:2016}.
%For the purposes of this work, the summarized traffic records will simply be referred to as `flow data'.
Recall that the ISCX IDS 2012 dataset is provided in PCAP as well as a custom XML file of network flows with associated ground-truth labels (indicating malicious or benign flows). The XML file is parsed and converted to a CSV file of flows, which becomes the input into the machine learning pipeline.
Recall that the CIC IDS 2017 dataset is in the form of both PCAP as well as flows
%records via the \textit{CICFlowMeter} tool.
%CICFlowMeter is a network traffic flow generator written in Java that takes raw PCAP as input and generates bidirectional flows where the first packet determines the forward (source to destination) and backward (destination to source) directions.
%Recall also that each flow is also accompanied or 
characterized by 74 usable features (5 categorical and 69 statistical).
%77\footnote{86? be consistent} statistical features. 
%Once a dataset is represented in the form of flows, it undergoes a process of cleaning and pre-processing to make sure that there are no erroneous characters in the data, removing fields that contain all zero or null values, and removing or modifying NaN (not a number) values. 
%(the features available for CIC IDS 2017 specifically are detailed in Table \ref{tab:cicids2017-features}), 
%such as duration, number of packets, number of bytes, length of packets, which are also calculated separately in the forward and backwards directions.
%Along with these statistical features for each flow, CICFlowMeter provides the corresponding source/destination IP address and port, protocol number, timestamp, label, and a unique FlowID for each flow record.

In order to make machine learning algorithms train models in the same feature space, it is a common practice to normalize or scale the continuous values among all the features. For this purpose, we use the standard \textit{min-max scaling}, which is a normalization method for scaling data to [0,1] as follows:
%In this work, Min-Max scaling is used to normalize the features during this step.
$X_{norm} = \frac{X - X_{min}}{X_{max} - X_{min}}$,
where $X_{min}$ and $X_{max}$ are respectively the minimum and maximum value of feature $X$.

In order to train DNNs over categorical data, we need to
%Therefore, to experiment with using these high dimensional categorical features, they must be 
convert them to numerical values. For this purpose, 
%there are a number of approaches.
%for accomplishing this, ranging from simple to more complex.
we propose adopting the \textit{entity embedding} technique \cite{guo2016entity} because 
%(i) embeddings can help models generalize better when data is sparse and statistics are unknown and (ii) 
it can cope with categorical features that take a large number of possible values. This is true for the datasets we analyze because
%\footnote{if we are talking about individual feature, why bother to consider the sum of their unique values in the previous subsection?} 
% NOTE: removed mention in previous section as not relevant
%is of high cardinality \cite{guo2016entity}.
%, which is the case of the two datasets we analyze.
%According to Cheng Guo and Felix Berkhan \cite{guo2016entity}, embeddings can help models to generalize better when data is sparse and statistics are unknown. They mention that this technique is especially useful for datasets which contain high cardinality features, where other methods would tend to overfit.
%For the datasets we analyze, thease are especially applicable because 
there are many possible values for source IP addresses, destination IP addresses, source port numbers, and destination port numbers.
%Therefore in this work, the embedding technique is used for these high cardinality features.
In the entity embedding method, the number of embedding dimensions are determined according to the following rule of thumb \cite{google2019mlcrashcourse}:
\begin{equation}
	dimensions = \ceil[\Big]{\sqrt[4]{possible\:values}\:},
	\label{eq:embedding}
\end{equation}
where $possible\:values$ is the number of possible values a categorical feature can take.
%Therefore, the high cardinality features for each dataset are reduced in dimension using entity embeddings,
%{\color{blue}
Specifically, a categorical feature is first mapped to an integer between $0$ and $n-1$, where $n$ is the number of unique values that can be taken by the feature,
%according to Eq. \eqref{eq:embedding} (i.e., the number of dimensions) 
and then encoded as a {\em dense} vector according to the dimensions as calculated in Eq. \eqref{eq:embedding}.
%}
%Eq. \eqref{eq:embedding}.
%When these categorical variables are embedded, they are then represented by a dense vector of floating point numbers according to the dimension that are calculated.
Table \ref{table:embedding-cicids} summarizes the embedding result of the four categorical features in the CIC IDS 2017 dataset.

% \begin{table}[htbp]
% \caption{Embedding Categorical Variables - ISCX IDS 2012 Dataset}
% \centering
% \label{table:embedding-iscx}
% \begin{tabular}{|c|c|c|}
% \hline
% \textbf{Feature}&\textbf{Possible Values}&\textbf{Embedded Dimensions} \\ \hline  
%  Source IP& 2,478 & 7  \\ \hline
%  Destination IP& 34,552 & 13  \\ \hline
%  Source Port& 64,482 & 16 \\ \hline 
%  Destination Port& 24,238 & 12 \\ \hline  
% \end{tabular}
% \end{table}

\begin{table}[!htbp]
\caption{Embedding of the four categorical features in the CIC IDS 2017 dataset.}
\centering
\label{table:embedding-cicids}
\begin{tabular}{|c|c|c|}
\hline
\textbf{Feature}&\textbf{Possible Values}&\textbf{Embedded Dimensions} \\ \hline  
 Source IP& 17,002 & 12  \\ \hline
 Destination IP& 19,112 & 12  \\ \hline
 Source Port& 64,638 & 16 \\ \hline 
 Destination Port& 53,791 & 15 \\ \hline  
\end{tabular}
\end{table}

% After running experiments and tweaking various hyperparameters, high performance for CIC IDS 2017 is achieved using a neural network configuration as shown in Figure \ref{fig:cicids-nn}. A similar structure is also used for ISCX IDS 2012.

% \begin{figure*}[!htbp]
% 	\centering
% 	\includegraphics[scale=0.3]{figures/cicids-dnn-structure.png}
% 	\caption{Deep Neural Network Architecture for CIC IDS 2017 Dataset}
% 	\label{fig:cicids-nn}
% \end{figure*}

The parameters (weights) for the vector representation of the categorical features are initialized using a random uniform distribution over the support $[-0.05, 0.05]$. %{\color{red}
This  representation is not only more computationally efficient, but the entity embedding layer learns intrinsic properties of each categorical feature,
%y (i.e., IP address, port), 
and the deeper layers of the neural network form complex combinations between them \cite{guo2016entity}.
%inherent relationship information between the various categorical variables, the other features in the dataset, and the label.
%}
%\footnote{need reference to justify this}
Since these vectors are inputs into the first layer of a neural network, their weights are updated in the back-propagation step at each epoch.

%After running experiments and tweaking various hyperparameters, high performance was achieved for both ISCX IDS 2012 and CIC IDS 2017 datasets.

%%%%%%%%%%%%%%%%%%%%%%%%%
%%%% CICIDS2017 RESULTS %%%
%%%%%%%%%%%%%%%%%%%%%%%%%
% \begin{table}[!htbp]
% \caption{CIC IDS 2017 Evaluation Results - Metrics}
% \label{table:cicids-eval-results}

% \noindent \centering{}%
% \begin{tabular}{|c|c|c|}
% \hline 
%  \textbf{Metric} & \textbf{With IPs} & \textbf{Without IPs} \tabularnewline
% \hline 
% True Negative & $749,299$ & $745,606$ \tabularnewline
% \hline 
% False Positive & $237$ & $3,930$ \tabularnewline
% \hline 
% False Negative & $137$ & $5,930$ \tabularnewline
% \hline 
% True Positive & $183,527$ & $177,734$ \tabularnewline
% \hline 
% Area Under the Curve & $0.9995$ & $0.9812$ \tabularnewline
% \hline 
% Accuracy & $0.9996$ & $0.9894$  \tabularnewline
% \hline 
% Error Rate & $0.0008$ & $0.0106$ \tabularnewline
% \hline 
% \textbf{True Positive Rate} & $\textbf{0.9993}$ & $\textbf{0.9677}$  \tabularnewline
% (Sensitivity, Recall, Detection Rate) &  &\tabularnewline
% \hline 
% True Negative Rate (Specificity) & $0.9997$ & $0.9948$ \tabularnewline
% \hline 
% \textbf{False Positive Rate} & $\textbf{0.0003}$ & $\textbf{0.0052}$  \tabularnewline
% \hline 
% \textbf{False Negative Rate} & $\textbf{0.0007}$ & $\textbf{0.0323}$ \tabularnewline
% \hline
% Precision & $0.9987$ & $0.9784$ \tabularnewline
% \hline 
% \textbf{F1 Measure} & $\textbf{0.9990}$ & $\textbf{0.9730}$  \tabularnewline
% \hline 
% \end{tabular}
% \end{table}

\subsubsection{Training}
The neural network consists of three layers of 64 units per layer.
Feeding into these three hidden layers is an initial input layer consisting of the embedded categorical variables concatenated with the statistical input features.
The activation function on each hidden layer is the ReLU activation function, $R(z) = max(0, z)$, while the last output layer uses a sigmoid activation function, $\sigma(z) = \frac{1}{1 + e^{-z}}$.
% given by Eq. \eqref{eq:relu} below:
% \begin{equation}
% 	R(z) = max(0, z).
% 	\label{eq:relu}
% \end{equation}
% The last output layer uses a sigmoid activation function given by Eq. \eqref{eq:sigmoid} below:
% \begin{equation}
% 	\sigma(z) = \frac{1}{1 + e^{-z}}.
% 	\label{eq:sigmoid}
% \end{equation}
A dropout rate of $0.40$ is used on each of the hidden layers.
The optimizer used is RMSProp, with a default learning rate of $0.001$.
%After initially using the Adam optimizer, it was found that RMSProp performed better at optimizing the loss function, enabling better results with validation and test sets.
The loss function used is binary crossentropy:
\begin{equation}
	H_p(q) = -\frac{1}{N} \sum_{i=1}^N y_i \cdot \log(p(y_i)) + (1 - y_i) \cdot \log(1 - p(y_i)),
	\label{eq:binary-crossentropy}
\end{equation}
where $y_i$ is the label ($1$ for malicious and $0$ for benign), $p(y_i)$ is the predicted probability of a given flow, and $N$ is the total number of flows. Intuitively, Eq. \ref{eq:binary-crossentropy} says that for each malicious flow ($y_i=1$), the loss is $\log(p(y_i))$, which is the logarithm of the probability that the flow is malicious;
for each benign flow ($y_i=0$), the loss is $\log(1-p(y_i))$, which is the logarithm of the probability that the flow is benign. 

%\footnote{describe the trained model: what hyperparameters; what optimization method, ...}

\subsubsection{Experiments and results}

%\footnote{i stop here: 1. the fonts on the x-axis and y-axis labels are too small--make them as large as the fontsize in the text; 2. merge the five figures into one figure such that each figure is a subfigure--using the s}

We aim to use experiments to answer two questions: (i) Is deep learning more effective than other machine learning methods? (ii) Is deep learning robust in the presence of dynamic IP addresses? Note that (ii) is important because a trained DNN, which uses IP addresses as an important feature, can easily become useless in the presence of dynamic IP addresses, which are produced by networks using the Dynamic Host Configuration Protocol (DHCP).

In order to answer the aforementioned question (i), we compare the effectiveness of deep learning and other machine learning methods using two standard metrics \cite{Pendleton16}, namely 
the {\em True-Positive Rate} (TPR) 
%or \textbf{Detection Rate} (DR): $TPR=DR=\frac{TP}{TP+FN}$ 
and the {\em False-Positive Rate} (FPR). %$FPR=\frac{FP}{FP+FN}$.

%{\color{red}....explain only the metrics you used, not copying from the paper ... also what average means here?}.
% \begin{itemize}
% \item \textbf{True Positives} (TP) are the number of samples that are \textit{correctly predicted as positive} (e.g. ground truth is `malicious' and the prediction is also `malicious').
% \item \textbf{True Negatives} (TN) are the number of samples that are \textit{correctly predicted as negative} (e.g. ground truth is `benign' and the prediction is also `benign').
% \item \textbf{False Positives} (FP) are the number of samples that are negative but predicted as positive (e.g. ground truth is `benign' and prediction is `malicious').
% \item \textbf{False Negatives} (FN) are the number of samples that are positive but are predicted as negative (e.g. ground truth is `malicious' and prediction is `benign').
% \end{itemize}

\ignore{
\begin{itemize}
    \item \textbf{True Positive Rate} (TPR) or \textbf{Detection Rate (DR)}: $TPR=DR=\frac{TP}{TP+FN}$
    \item \textbf{True Negative Rate} (TNR): $TNR=\frac{TN}{TN+FP}$.
    \item \textbf{False Positive Rate} (FPR): $FPR=\frac{FP}{FP+FN}$
    \item \textbf{False Negative Rate} (FNR): $FNR=\frac{FN}{FN+TP}$.
\end{itemize}
}

%%%%%%%%%%%%%%%%%%%%%%%%%%%%%%%%%%%%
%%%% CICIDS2017 COMPARISON TABLE %%%
%%%%%%%%%%%%%%%%%%%%%%%%%%%%%%%%%%%%
\begin{table}[!htbp]
\caption{Comparison of deep learning based intrusion detection and other machine learning methods based intrusion detection \cite{ahmim2018novel} using the CIC IDS 2017 dataset.}
\label{table:cicids-result-comparison}

\noindent \centering{}%
\begin{tabular}{|c|c|c|}
\hline 
 \textbf{Technique} & \textbf{TPR} & \textbf{FPR}\tabularnewline
\hline 
\makecell{Hypbrid IDS \\ Decision Tree + Rule-based} \cite{ahmim2018novel} & $0.94475$ & $.01145$ \tabularnewline
\hline 
WISARD \cite{de2018experimental} & $0.48175$ & $0.02865$ \tabularnewline
\hline 
Forest PA \cite{adnan2017forest} & $.92920$ & $0.03550$  \tabularnewline
\hline 
J48 Consolidated \cite{ibarguren2015coverage} & $0.92020$ & $0.06645$ \tabularnewline
\hline 
LIBSVM \cite{chang2001libsvm} & $0.54595$ & $0.05130$ \tabularnewline
\hline 
FURIA \cite{huhn2009furia} & $0.90500$ & $0.03165$  \tabularnewline
\hline 
Random Forest \cite{ahmim2018novel} & $0.93050$ & $0.01880$  \tabularnewline
\hline 
REP Tree \cite{ahmim2018novel} & $0.91640$ & $0.04835$   \tabularnewline
\hline 
MLP \cite{ahmim2018novel} & $0.77830$ & $0.07350$  \tabularnewline
\hline 
Naive Bayes \cite{ahmim2018novel} & $0.82510$ & $0.33455$ \tabularnewline
\hline 
Jrip \cite{ahmim2018novel} & $0.93400$ & $0.04470$ \tabularnewline
\hline 
J48 \cite{ahmim2018novel} & $0.91990$ & $0.05040$ \tabularnewline
\hline 
\textbf{DNN with IPs} & $\textbf{0.9993}$ & $\textbf{0.0003}$   \tabularnewline
\hline 
\textbf{DNN without IPs} & $\textbf{0.9677}$ & $\textbf{0.0052}$  \tabularnewline
\hline
\end{tabular}
\end{table}

Table \ref{table:cicids-result-comparison} compares deep learning against the other approaches evaluated in \cite{ahmim2018novel} using the CIC IDS 2017 dataset.
We observe that DNN while using IP addresses leads to the highest True-Positive Rate (Detection Rate) and lowest False-Positive Rate.
%used in this study outperform other approaches.
This leads to the following:

\begin{insight}
%Embedding high-cardinality categorical features of IP address and port numbers 
DNN while using IP addresses achieves the highest effectiveness when compared with the other machine learning method studied in the literature.
%    The neural network forms a memory and learns inherent relationships amongst itself, other features, and the label.
	%When using the IP address as a feature and representing it as a dense embedded vector at the first layer of the deep neural network, the best results are achieved.
	%It is theorized that when using the embedding technique, the neural network forms a type of memory of the IP address in relation to the other features and the label.%, whereby it enables the highest performance for the classification task.
	%It should be noted, however, that the IP address feature is highly relevant to the given dataset from which it originates, and may therefore not generalize well to other datasets.
\end{insight}

\begin{figure*}[!htbp]
    \begin{subfigure}[b]{0.19\textwidth}
    	\centering
    	\includegraphics[width=\textwidth]{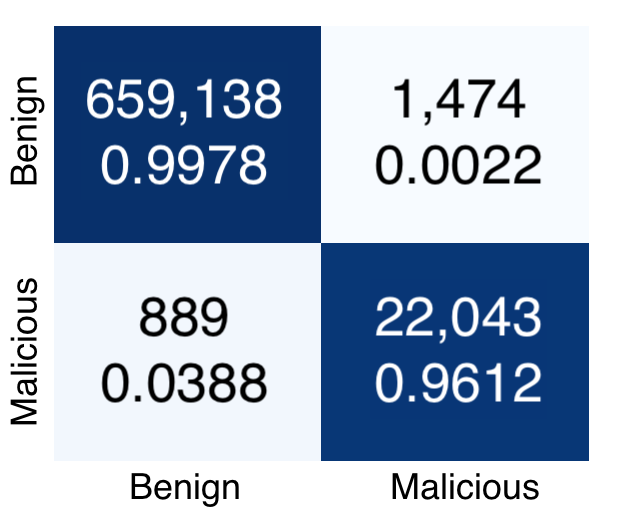}
    	\caption{ISCX2012 w/ IP}
    	\label{fig:iscx-confusion-matrix-embeddings}
    \end{subfigure}
%    \hfill
    \begin{subfigure}[b]{0.19\textwidth}
    	\centering
    	\includegraphics[width=\textwidth]{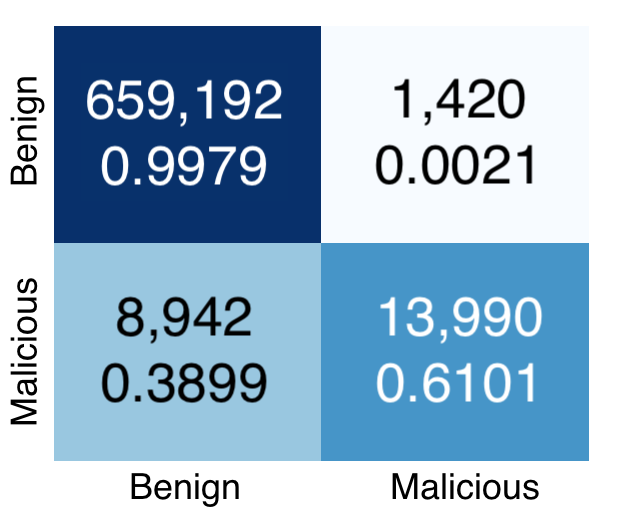}
    	\caption{ISCX2012 w/o IP}
    	\label{fig:iscx-confusion-matrix-noips}
    \end{subfigure}
    \begin{subfigure}[b]{0.19\textwidth}
    	\centering
    	\includegraphics[width=\textwidth]{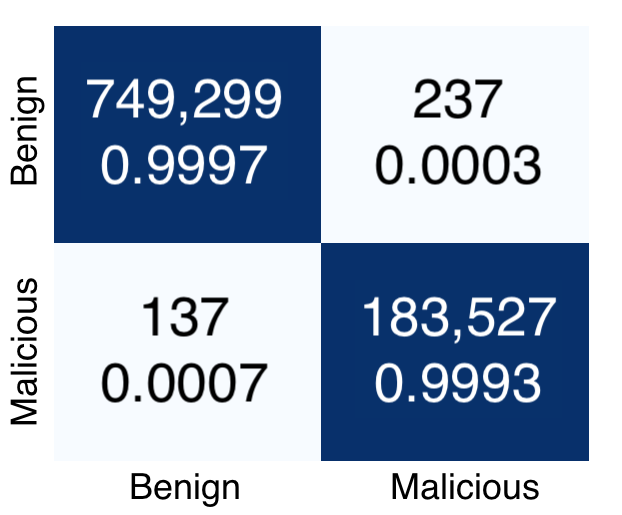}
    	\caption{CIC2017 w/ IP}
    	\label{fig:cicids-confusion-matrix-embeddings}
    \end{subfigure}
%    \hfill
    \begin{subfigure}[b]{0.19\textwidth}
    	\centering
    	\includegraphics[width=\textwidth]{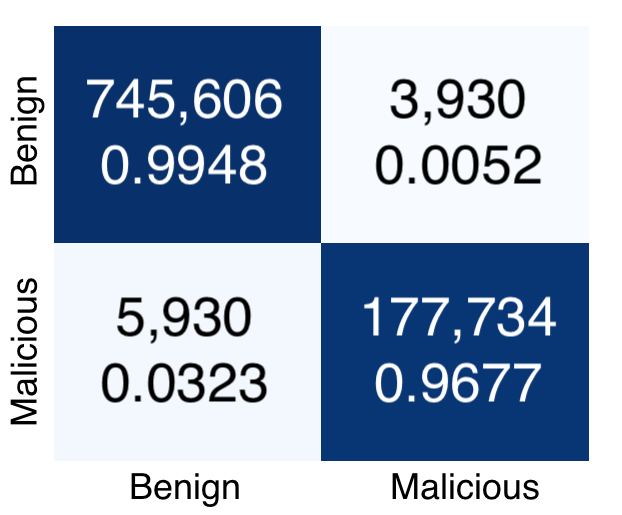}
    	\caption{CIC2017 w/o IP}
    	\label{fig:cicids-confusion-matrix-embeddings-no-ips}
    \end{subfigure}
    \begin{subfigure}[b]{0.19\textwidth}
    	\centering
    	\includegraphics[width=1.10\textwidth]{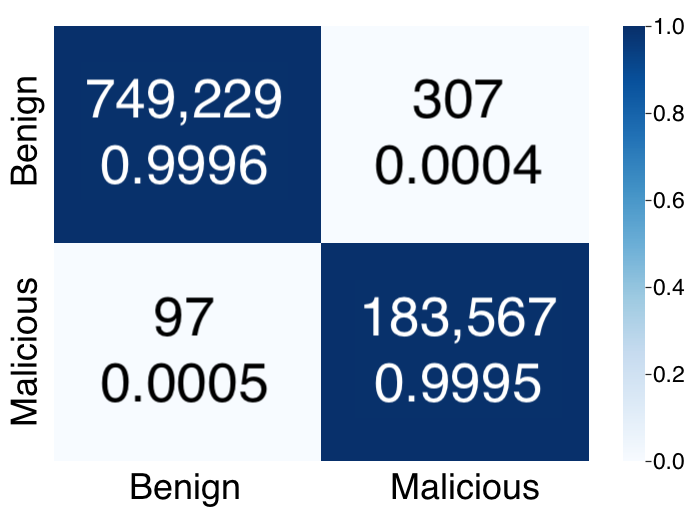}
    	\caption{CIC2017 - first 3 octets}
    	\label{fig:cicids-confusion-matrix-first-3-octets}
    \end{subfigure}
    \caption{Confusion matrix results for both datasets, where the $x$-axis is the predicted class and the $y$-axis is the true class.
    %{\color{red}the fonts on the labels are too small; enlarge them; possibly draw only one legend bar because they appear to be the same}
    }
    \label{fig:cf}
\end{figure*}

In order to answer the aforementioned question (ii), we train the deep learning model using some portion of the IP addresses. This is reasonable because DHCP typically operates in the same network, meaning that the network identity is static (e.g., the first 24-bit of IP addresses of a class C network).

% \begin{insight}
%     When including Source/Destination IP Address and Ports, neural network performs best at classifying malicious and benign flows.
%     For CIC IDS 2017, in experiments that omit the Source/Destination IP address features, but still utilizes detailed flow statistics, the neural network performs nearly as well, achieving an F1 Measure of $0.9730$ in comparison to the experiment that included the IP address features, which achieved the $0.9980$ F1 score.
%     For ISCX IDS 2012, when Source/Destination IP address features are not included, the performance drops signifanctly going from a detection rate of $96.12\%$ down to $61.01\%$.
%     Therefore, training a neural network by omitting IP address features, but ensuring it contains detailed flow statistics, may be a viable way to use the trained model on a different computer network it has not seen before.
% \end{insight}
%\footnote{Move the relevant insights up here ...}

%\footnote{revise the following ... in the style i put above ...}
Figure \ref{fig:cf} shows the results for the two datasets with and without IP address features. 
%{\color{red}explain the subfigures one-byone...
Figure \ref{fig:cicids-confusion-matrix-first-3-octets} shows the results of using just the first three octets for source and destination IP address. 
%{\color{red} highlight observations from the picture ... otherwise why bother to put there ... revise the insight and the next subsection similarly }
In Figure \ref{fig:iscx-confusion-matrix-noips} we observe that for the ISCX IDS 2012 dataset, when removing the IP address feature the performance drops considerably in terms of TPR and FNR.
For the CIC IDS 2017 dataset, Figure \ref{fig:cicids-confusion-matrix-embeddings-no-ips} shows that removing the IP address only slightly degrades the performance in comparison to ISCX IDS 2012.
Note that there is considerably larger amount of malicious examples in CIC IDS 2017 ($19.68\%$) compared to ISCX IDS 2012 ($3.32\%$).
We also notice that embedding the IP address with only the first three octets (Figure \ref{fig:cicids-confusion-matrix-first-3-octets}) achieves similar results as when using the full IP address as shown (Figure \ref{fig:cicids-confusion-matrix-embeddings}).
%}
%\footnote{make sure the discussion here has causal relation to the insight that is drawn below}
This leads to the following:

\begin{insight}
DNN while using the first three octets of the IP address is as effective as using the full IP address, meaning that deep learning based intrusion detection is robust in the presence of DHCP.
%	This adds robustness to the approach, since the last octet is more dynamic, while they first three tend to be more static.
	%This naturally captures interactions between different local subnets, and between local and remote hosts.
	%This methodology can be used to reliably include IP address as feature and overcome limitations imposed by DHCP.
However, using full IP addresses is important when the dataset is imbalanced (i.e., the proportion of labeled malicious traffic is small). 
\end{insight}

\subsection{Using Autoencoders for network intrusion detection}

% The configuration of the autoencoder used in this case study is shown in Figure \ref{fig:cicids-autoencoder-configuration}.

% \begin{figure}[!htbp]
% 	\centering
% 	\includegraphics[scale=0.3]{figures/autoencoder-structure.png}
% 	\caption{CIC IDS 2017 Autoencoder Configuration}
% 	\label{fig:cicids-autoencoder-configuration}
% \end{figure}

%\footnote{divide this subsection into subsubsection in a fashion similar to the previous subsection}

\subsubsection{Pre-processing}
For the Autoencoder experiments, all 69 usable continuous features of flow statistics in the CIC IDS 2017 dataset are used, and are normalized using the \textit{min-max} technique mentioned above. 
One categorical feature of ``protocol'' is also used, which only has 3 unique values, and is converted to floating point numbers using one-hot encoding.
The high-cardinality features of IP address and port are not used; we leave it to future work to incorporate these into the training of autoencoders.

\subsubsection{Training}

%The configuration of the autoencoder used in this study is shown in Figure \ref{fig:cicids-autoencoder-configuration}.
\ignore{
\begin{figure}[!htbp]
	\centering
	\includegraphics[scale=0.3]{figures/autoencoder-structure.png}
	\caption{CIC IDS 2017 Autoencoder Configuration {\color{red}font size problem ....}}
	\label{fig:cicids-autoencoder-configuration}
\end{figure}
}
The autoencoder configuration consists of 7 layers, with the first and last layer using the sigmoid activation function, and all other hidden layers using ReLU.
The first and last layer consist of 72 units (representing all input features), and the hidden layers consist of 140, 35, 16, 16, 35 units respectively.
In addition, L1 regularization is applied to the first input layer.
The objective function for the autoencoder is the squared error. Written out in terms of weights and inputs, this function is shown in Eq. \eqref{eq:squared-error} below.
\begin{equation} \label{eq:squared-error}
	J = | X - \hat{X} |_F^2 = | X - sigmoid(sigmoid(X \cdot W)W^T) |_F^2 
\end{equation}

\subsubsection{Experiment and results}

Figure \ref{fig:cicids-autoencoder-reconstruction-error} plots the experimental results. We observe that there is a higher reconstruction error for the malicious traffic flows as compared to the benign flows.
Depending on the threshold set, the number of false positives can be adjusted.
With the current threshold set at a $0.03$ reconstruction error, there only results in a total of $89$ false positives and a False-Positive Rate of $0.00013$.
However, there is a high False-Negative Rate of $0.7670$.
In addition, we observe that a majority of the malicious flows are clustered in groups, lending credence to future work that can incorporate the time domain as a feature.
%\footnote{is this all the interesting phenomenon the figure shows?}
%Future work should look at including categorical variables within the autoencoder to improve its detection rate.
%In terms of anomaly detection, it is ideal to have a low number of false positives.
We draw the following insight:

\begin{figure}[!htbp]
	\centering
	\includegraphics[width=.4\textwidth]{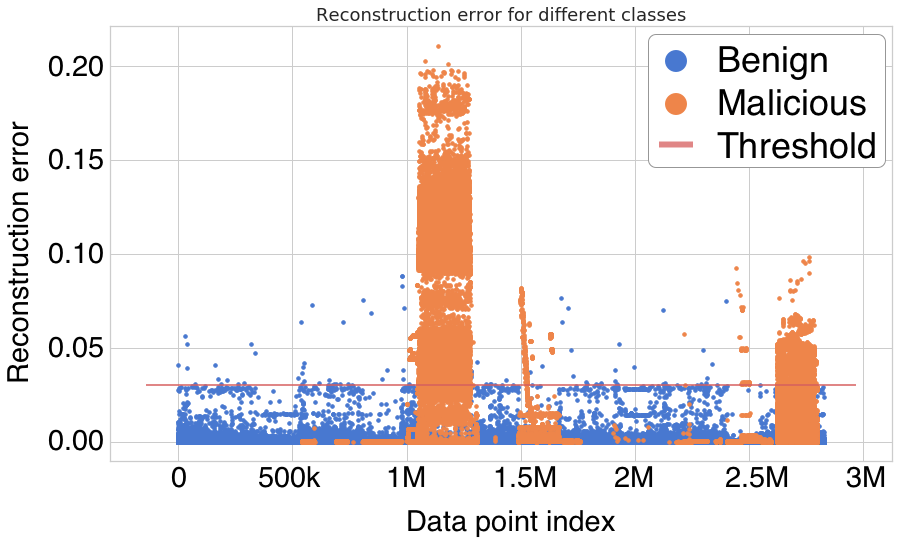}
	\caption{Experimental results using the CIC IDS 2017 dataset: Autoencoder reconstruction error with threshold.} 
	\label{fig:cicids-autoencoder-reconstruction-error}
\end{figure}

\begin{insight}
Autoencoders can be effective as anomaly detection mechanisms for network intrusion detection (in terms of low False-Positive Rate) when training on benign traffic only.
%, able to achieve a low false positive rate.
%    However, inclusion of IP address and port features, as well as other types of autoencoders (i.e., variational) need to be evaluated in an effort to reduce the high false negative rate.
    %They can be used to detect malicious flows that have never been encountered before with a low false positive rate.
    %Since this approach exhibits a low false positive rate, it can be useful for bringing the anomalies to the attention of a human analyst for review.
    %However, this it at the cost of a high false negative rate, and should be used along with other supervised techniques as part of a defense-in-depth strategy.
\end{insight}

%\subsection{Findings}

%list the important insights and discuss their real-world implications 

% \begin{insight}
% When using embeddings for the categorical variables of Source/Destination IP Address and Ports, neural network performs exceptionally well at classifying malicious and benign flows.
% \end{insight}

\section{Limitations}
\label{sec:limitations}

The present study has some limitations. From a methodology point of view, we only considered two kinds of neural networks. Future research needs to consider additional types of neural networks. 
%{\color{blue}
For DNNs, we need to conduct further experiments using only the first two octets (first 16 bits), or even the first one octet, to see if they can achieve the same results as using the full 32-bit IP address. 
For Autoencoders, we need to investigate whether or not using the IP address and port features 
%as well as other types of autoencoders (i.e., variational) need to be evaluated in an effort to 
can reduce their False-Negative Rate.
%}
From a datasets point of view, the ISCX IDS 2012 dataset 
%has the drawback that no consistent format of ground truth is provided.
%This dataset has some limitations: (i) 
contains only binary ground-truth labels (i.e., malicious vs. benign) and
%(ii) lacks a CSV version of flow records which can be consistently used by other studies;
%{\color{red}it only comes with attack vs. benign traffic, meaning that it can only be used for binary classification};
%as opposed to just a binary classifier in the case of the labeled ISCX IDS 2012 dataset.
contains no HTTP traffic.

\section{Conclusion}
\label{sec:conclusion}

%Deep neural networks can be used for both supervised and unsupervised learning in the field of network intrusion detection.
We have shown that DNN can achieve excellent results in supervised network intrusion detection.
%, given enough training examples.
%Using embeddings for the high dimensional categorical variables of source and destination IP addresses and ports, the neural network performs exceptionally well at classifying malicious and benign flows using supervised learning.
We also showed that using only the first three octets of IP addresses can be effective in coping with the use of dynamic IP addresses, indicating robustness of DNN in the presence of DHCP.
%because we can achieve the same network intrusion detection performance as using full IP addresses.
We further showed that autoencoders can be used for anomaly detection when they are trained on benign flows.
%, enabling the discovery of never-before-seen malicious flows with a low false positive rate.
%The limitations of the present study motivate future research directions.
%Future work should evaluate the most recent CSE-CIC-IDS2018 dataset in comparison with CIC IDS 2017 -- training a deep neural network on one computer network traffic environment, and determining how well it generalizes to classifying flows on an entirely different environment will be a good next step.

\noindent{\bf Acknowledgements}.  This work was supported in part by ARL grant \#W911NF-17-2-0127 and NSF CREST Grant \#1736209.

\bibliographystyle{ieeetr}
\bibliography{reference,metrics} % Entries are in the "refs.bib" file

%\vspace{12pt}

\end{document}